# The Coherence Field in the Field Perturbation Theory of Superconductivity.

Moshe Dayan

Department of Physics, Ben-Gurion University,
Beer-Sheva 84105, Israel.

E-mail: mdayan@bgu.ac.il

Key Words: Gorkov-Nambu field theory of superconductivity, High Temperature Superconductivity.

# The Coherence Field in the Field Perturbation Theory of Superconductivity

Moshe Dayan

Department of Physics, Ben-Gurion University,
Beer-Sheva 84105, Israel.

Abstract

We re-examine the Nambu-Gorkov perturbation theory of superconductivity on the basis of the Bogoliubov-Valatin quasi-particles. We show that two different fields (and two additional analogous fields) may be constructed, and that the Nambu field is only one of them. For the other field- the coherence field- the interaction is given by means of two interaction vertices that are based on the Pauli matrices $\tau_1$ and $\tau_3$. Consequently, the Hartree integral for the off-diagonal pairing self-energy may be finite, and in some cases large. We interpret the results in terms of conventional superconductivity, and also discuss briefly possible implications to HTSC.

## 1. INTRODUCTION

Recently, I have defined the Gorkov-Nambu formalism for the superconductive and pseudogap double correlations in HTSC [1]. In this recent analysis, as well as in a former one, which was solely devoted to the analysis of pseudogaps [2], Hartree integrals make the major parts of the superconductive order parameter, and the pseudo-order parameter. This is a key feature of the analysis, a feature that is evidently distinctive from the Nambu-Gorkov theory of conventional superconductivity. Technically, the Hartree integrals of the off-diagonal self-energies (and propagators) could be made significant only by allowing scattering vertices to be spanned by the same Pauli matrices that span the off-diagonal self-energies. This is so because the Hartree Feynman diagrams for the $\tau_j$ ($\tau_j = \tau_1, \tau_3$) self-energies, scale with the traces of $\tau_j m_i$, where $m_i$ is the matrix that define the interaction vertex, and the traces vanish unless $m_i = \tau_j$. It has been commented that interaction through the vertex $\tau_1$ does not conserve the particle number, but this feature could be justified when operating in the superconductive phase, which does not conserve the number operator anyway.

The need to study the problem of the off-diagonal interaction vertices in the superconductive state has motivated the present author to study the fundamentals of the Gorkov-Nambu theory in order to check its compatibility with such vertices. To go back to the fundamental Gorkov-Nambu (GN) theory, we assume conventional superconductivity, without the complications of the pseudogaps that exist in HTSC. Conjectured implications to HTSC will be discussed qualitatively only in the conclusion section. The GN theory is essentially a Hartree-Fock theory where the Fock potential is generalized to include "Cooper-pairing potential". This was assumed both by Gorkov [3], and by Nambu [4]. Although these two works are quite equivalent, Nambu's is presented more as a perturbation analysis. It assumes the field operator

$$\Psi_k^N = \begin{pmatrix} c_{k\uparrow} \\ c_{-k\downarrow}^+ \end{pmatrix}, \qquad (1)$$

and the unperturbed Hamiltonian

$$H_0 = \sum_k \{\Psi_k^{N+}[(\varepsilon_{0k} + \chi_k)\tau_3 + \Delta_k\tau_1]\Psi_k^N + \varepsilon_k\}, \tag{2}$$

where $\varepsilon_{0k}$ is the band excitation energy, $\Delta_k$ is the Fock pairing potential, $\chi_k$ is the regular Hartree-Fock potential, and $\varepsilon_k = \varepsilon_{0k} + \chi_k$. The Hamiltonian is the sum of the unperturbed Hamiltonian and the interaction Hamiltonian, $H = H_0 + H_i$. Since the Hartree-Fock potentials are added to the unperturbed Hamiltonian, they should be subtracted form the interaction Hamiltonian, which consequently is given by

$$H_i = \frac{1}{2}\sum_{k,k',q} V_q \{\Psi_{k+q}^{N+}\tau_3\Psi_k^N\}\{\Psi_{k'-q}^{N+}\tau_3\Psi_{k'}^N\} - \sum_k \Psi_k^{N+}[\chi_k\tau_3 + \Delta_k\tau_1]\Psi_k^N. \tag{3}$$

The off-diagonal potential in the GN formalism is actually a Fock potential. Only the diagonal potential $\chi_k$ may have a Hartree contribution. Since the Hartree-Fock potentials are subtracted, the self-energy vanishes in the Hartree-Fock approximation. In particular, the off-diagonal self-energy that results from the Feynman-Dyson perturbation serious vanishes, which self-consistently results in the well-known integrals for the pairing self-energy [4,5].

One necessary condition for the application of the Wick's theorem, which is essential for the validity of the Feynman-Dyson perturbation serious, is that it is based on the appropriate excitations. The mathematical implication of this condition is that the annihilation operators that construct the field should yield zero when applied on the ground state. A direct test of applying the GN field operator on the superconductive ground state

$$|\Phi_0> = \prod_k |\Phi_k> = \prod_k (u_k + v_k c_{k\uparrow}^+ c_{-k\downarrow}^+)|0> \tag{4}$$

yields $\Psi_k^N |\Phi_0> = \prod_{k'\neq k}|\Phi_{k'}> \begin{pmatrix} v_k c_{-k\downarrow}^+ \\ u_k c_{-k\downarrow}^+ \end{pmatrix}|0> \neq 0$, which does not vanish. However, in the following we shall show that the GN operator is in agreement with Wick's

theorem. It is one of four fields operators, which make two analogous groups, that are in accord with Wick's theorem.

The "natural" elementary excitations of the superconductive state, that any field should be based on, are undoubtedly the Bogoliubov-Valatin excitations [6,7]. Therefore, in the present paper, we define from the start field operators that are based on the Bogoliubov-Valatin (BV) excitations, and construct superconductive fields from these operators. It turns out that two different fields could be constructed from these operators, plus two more analogous fields. By "analogous fields" we mean fields that are equivalent in all respects, except for the sign of the order parameter. All these four fields are compatible with Wick's theorem, and with experiment. The GN field is just one of these four fields. Together with its analog, it makes the GN field pair. The other pair yields the same experimental results in regular superconductors, despite some essential differences. One such important difference is the interaction vertices that are spanned by the $\tau_1$ Pauli matrix. This in turn, suggests the validity of the off-diagonal Hartree diagram, which is the main motivation to the present analysis. We show that in conventional superconductors this Hartree diagram vanishes, in accordance with the success of the GN theory for conventional superconductivity. However, we conjecture that in HTSC the Hartree diagram may be the major source of the superconductive energy gap.

2. THE FIELDS OF THE BOGOLIUBOV-VALATIN EXCITATIONS.

The superconductive state cannot evolve from the normal state by any perturbation theory. It must be based on the superconductive ground state, and on the superconducting field. The latter must be based on the elementary excitations of the superconductive state, which were proposed by Bogoliubov and (independently) by Valatin to be

$$\gamma_k = u_k c_{k\uparrow} - v_k c^+_{-k\downarrow}, \tag{5a}$$

$$\gamma_{-k} = \eta_k = u_k c_{-k\downarrow} + v_k c^+_{k\uparrow}. \tag{5b}$$

where $u_k$ and $v_k$ are the known BCS coherence parameters, and the c's are the known electronic operators in the normal state. The Bogoliubov-Valatin (BV) operators diagonalize the unperturbed Hamiltonian of Eq. (3), and yield the following relations

$$\gamma_k |\Phi_0> = \eta_k |\Phi_0> = 0, \tag{6}$$

$$\{\gamma_k, \gamma_{k'}^+\} = \{\eta_k, \eta_{k'}^+\} = \delta_{k,k'}, \tag{7}$$

$$\{\gamma_k, \gamma_{k'}\} = \{\eta_k, \eta_{k'}\} = \{\gamma_k, \eta_{k'}^+\} = \{\gamma_k^+, \eta_{k'}\} = 0. \tag{8}$$

Equations (6-8) guarantee the validity of Wick's theorem, provided that the field operators are constructed by means of the BV operators.

Let us propose the following superconductive vector field

$$\Psi_k(t) = \begin{pmatrix} u_k \\ v_k \end{pmatrix} \gamma_k(t) + \begin{pmatrix} -v_k \\ u_k \end{pmatrix} \eta_k^+(t), \tag{9a}$$

where all the operators are in the interaction picture. The field operator may also be written in relation to $\Psi_k^N$ as

$$\Psi_k(t) = \frac{1}{E_k} \begin{bmatrix} \varepsilon_k & -\Delta_k \\ \Delta_k & \varepsilon_k \end{bmatrix} \Psi_k^N(t), \tag{9b}$$

where $E_k = \sqrt{\varepsilon_k^2 + \Delta_k^2}$, We see that in the limit of $\Delta_k \to 0$, we get $\Psi_k \to sign(\varepsilon_k) \Psi_k^N$. In the spatial space we have

$$\Psi(x,t) = \int \frac{d^3k}{(2\pi)^3} \Psi_k(t) \exp(i\mathbf{k} \cdot \mathbf{x}). \tag{10}$$

The unperturbed Hamiltonian is given by

$$H_0 = i \int \frac{d^3x}{(2\pi)^3} \Psi^+(x,t) \frac{d}{dt} \Psi(x,t), \tag{11}$$

$$H_0 = \sum_k E_k (\gamma_k^+ \gamma_k - \eta_k \eta_k^+) = \sum_k E_k (\gamma_k^+ \gamma_k + \eta_k^+ \eta_k - 2v_k^2). \tag{12}$$

The Hamiltonian of Eq. (12) is identical with the BV Hamiltonian (except possibly for a constant term, which does not affect the discussed physics) [8,9]. Note that the excitation energy in the normal state is $\varepsilon_k = \varepsilon_{0k} + \chi_k$, where the contributions to $\chi_k$ come from the Hartree integral as well as the Fock integral. However, since the Hartree integral is independent of energy, it is compensated by a shift of the Fermi level, leaving only the Fock integral. The pairing potential $\Delta_k$ is by definition only a Fock potential.

The unperturbed matrix Green's function is

$$G_0(k,t) = -i <\Phi_0 | T\{\Psi(k,t), \Psi^+(k,0)\} | \Phi_0>$$

$$= -i <\Phi_0 | T\{U_k \gamma_k(t)\gamma_k^+(0) + W_k \eta_k^+(t)\eta_k(0)\} | \Phi_0> \tag{13a}$$

$$= -i <\Phi_0 | \{U_k \exp(-iE_k t)\Theta(t) - W_k \exp(iE_k t)\Theta(-t)\} | \Phi_0>, \tag{13b}$$

with

$$U_k = \begin{bmatrix} u_k^2 & u_k v_k \\ u_k v_k & v_k^2 \end{bmatrix}, \text{ and } W_k = \begin{bmatrix} v_k^2 & -u_k v_k \\ -u_k v_k & u_k^2 \end{bmatrix}. \tag{13c}$$

The time Fourier transform of Eqs. (13b) yields

$$G_{011}(\omega,k) = \frac{u_k^2}{\omega - E_k + i\delta} + \frac{v_k^2}{\omega + E_k - i\delta} = \frac{\omega + \varepsilon_k}{\omega^2 - E_k^2 + i\delta}, \tag{14a}$$

$$G_{022}(\omega,k) = \frac{v_k^2}{\omega - E_k + i\delta} + \frac{u_k^2}{\omega + E_k - i\delta} = \frac{\omega - \varepsilon_k}{\omega^2 - E_k^2 + i\delta}, \tag{14b}$$

$$G_{012}(\omega,k) = G_{021}(\omega,k) = \frac{u_k v_k}{\omega - E_k + i\delta} - \frac{u_k v_k}{\omega + E_k - i\delta} = \frac{\Delta_k}{\omega^2 - E_k^2 + i\delta}, \tag{14c}$$

where the well-known BCS relations between $\varepsilon_k$, $\Delta_k$, and the coherence parameters have been applied. The unperturbed propagators may be written in a matrix form as

$$G_0(\omega,k) = \frac{\omega I + \varepsilon_k \tau_3 + \Delta_k \tau_1}{\omega^2 - E_k^2 + i\delta}. \tag{15}$$

The propagators in Eqs. (14) and (15) are in agreement with the GN theory [4,5]. Features of disagreement will be discussed in the next sections. The Hamiltonian of Eq. (12) and the form of the propagators of Eqs. (14) and (15) confirm the validity of the field of Eq. (9), as the proper superconductive field.

The obtainment of Eqs. (12), (14) and (15) could be achieved from other fields which yields the same matrices $U_k$ and $W_k$. One of these fields is

$$\overline{\Psi}_k^N(t) = \begin{pmatrix} u_k \\ v_k \end{pmatrix} \gamma_k(t) + \begin{pmatrix} v_k \\ -u_k \end{pmatrix} \eta_k^+(t) = \tau_3 \Psi_k^N(t). \tag{16}$$

This field is the analogous of the GN field. The GN field itself, by means of the BV operators, is given by

$$\Psi_k^N(t) = \begin{pmatrix} u_k \\ -v_k \end{pmatrix} \gamma_k(t) + \begin{pmatrix} v_k \\ u_k \end{pmatrix} \eta_k^+(t). \tag{17}$$

Note that the off-diagonal elements of $U_k$ and $W_k$, of the fields $\Psi_k^N$ and $\overline{\Psi}_k^N$, are interchanged. Consequently, for the GN field $\Delta_k = -2u_k v_k E_k$, whereas for $\overline{\Psi}_k^N$ we have $\Delta_k = 2u_k v_k E_k$. The two fields are analogous. The field analogous to $\Psi_k(t)$ of Eq. (9a) is given by

$$\overline{\Psi}_k(t) = \begin{pmatrix} -u_k \\ v_k \end{pmatrix} \gamma_k(t) + \begin{pmatrix} v_k \\ u_k \end{pmatrix} \eta_k^+(t) = -\frac{1}{E_k} \begin{bmatrix} \varepsilon_k & \Delta_k \\ \Delta_k & -\varepsilon_k \end{bmatrix} \Psi_k^N. \tag{18}$$

We see that the two pairs of analogous fields are obtained by assigning a minus sign to a different coherence parameter (in Eq. (9a), and in Eqs. (16-18)) for each of the four fields. Each one of the four fields is a vector linear combination of the normal state operators $c_{k\uparrow}$, and $c_{-k\downarrow}^+$. In the GN pair there is a complete destructive interference of the operator $c_{-k\downarrow}^+$ in the upper component, and of the operator $c_{k\uparrow}$, in the lower component. In the other pair both operators are mixed in each component, giving rise to the coherence factors that we shall see in the next section. Therefore, these fields will be referred to as the "coherence fields". Analogous fields have opposite signs for their energy gaps, for one $\Delta_k = 2u_k v_k E_k$, while for its analog $\Delta_k = -2u_k v_k E_k$. In BCS, the parameters $u_k$ and $v_k$ are obtained by equations for their squares. Their sign is arbitrary, but for the sake of simplicity we assume real and positive $u_k$ and $v_k$. The sign of $\Delta_k$ is also arbitrary, since in experimental functions the quantity $\Delta_k^2$ appears. In equations where $\Delta_k$ appears (not squared), the arbitrary sign might always be compensated by another arbitrary sign (such as that of $v_k$, or of another energy gap). This is a source of the sign-duality that exists for each pair of analogous fields. However, the arbitrariness of the gap sign is applicable for the field, but cannot be manipulated once the field has been chosen.

## 3. THE INTERACTION HAMILTONIAN AND THE PAIRING SELF-ENERGY.

The interaction Hamiltonian of the GN theory is given by Eq. (3). The interaction vertices are bilinear forms of the c-operators on the Pauli matrix $\tau_3$. The Hartree diagram for the off-diagonal self-energy is finite only when we allow interaction vertices to be spanned by $\tau_1$. Within the basis of the c-operators, any $\tau_1$ interaction vertex is a product of two creation (or two annihilation) operators, which is forbidden by gauge invariance. The same holds for the analogous field $\overline{\Psi}_k^N$. In Ref.[1], I have justified $\tau_1$ interaction vertices, because the superconductive state is not gauge-invariant any way. Although this argument is correct, its application with respect to the field $\Psi_k^N$ is questionable. Here we show that the $\tau_1$ interaction vertices emerge naturally for the coherence field-$\Psi_k$.

To express the interaction Hamiltonian of Eq. (3) by means of the field $\Psi_k(t)$, we write the inverse of Eq. (9b)

$$\Psi_k^N(t) = \frac{1}{E_k}\begin{bmatrix} \varepsilon_k & \Delta_k \\ -\Delta_k & \varepsilon_k \end{bmatrix}\Psi_k(t) = M_k \Psi_k(t). \tag{19}$$

Each interaction vertex in the first term of the interaction Hamiltonian becomes

$$Int.Vertex(k_1,k_2,t) = \frac{\sqrt{V_q}}{E_1 E_2}\Psi_{k_1}^+(t)\begin{bmatrix} \varepsilon_1\varepsilon_2 - \Delta_1\Delta_2 & \varepsilon_1\Delta_2 + \Delta_1\varepsilon_2 \\ \varepsilon_1\Delta_2 + \Delta_1\varepsilon_2 & -(\varepsilon_1\varepsilon_2 - \Delta_1\Delta_2) \end{bmatrix}\Psi_{k_2}(t), \tag{20a}$$

$$= \sqrt{V_q}\Psi_{k_1}^+(t)I(k_1,k_2)\Psi_{k_2}(t), \tag{20b}$$

where $I(k_1,k_2) = M_{k_1}^{-1}\tau_3 M_{k_2}$, and for simple notation, the momenta $k_1$ and $k_2$ have been replaced by their subscripts 1 and 2. The matrix $I(k_1,k_2)$ is the matrix for the scattering vertex between $k_1$ and $k_2$,

$$I(k_1,k_2) = \frac{1}{E_1 E_2}[(\varepsilon_1\varepsilon_2 - \Delta_1\Delta_2)\tau_3 + (\varepsilon_1\Delta_2 + \Delta_1\varepsilon_2)\tau_1]. \tag{21}$$

Eq. (21) shows clearly that interaction vertex via $\tau_1$ emerges naturally for the coherence field $\Psi_k$. The equation provides simple prescription for the interactions in the $\Psi_k$ field: 1) Each scattering vertex must be replaced by two kinds of vertices, one via $\tau_3$, and the other one via $\tau_1$. 2) The usual scatterings potential $V_q$ may be used, provided that each $\tau_3$-vertex is factored by $(\varepsilon_1\varepsilon_2 - \Delta_1\Delta_2)$, and each $\tau_1$-vertex is factored by $(\varepsilon_1\Delta_2 + \Delta_1\varepsilon_2)$. We immediately recognize these two factors as the coherence factors that appear in the response of some experimental functions (such as in the acoustic attenuation rate [5]). When transforming the interaction Hamiltonian of Eq. (3) in terms of the coherence field $\Psi_k$, we notice that $\Psi_k$ and $\Psi_k^N$ correspond to opposite signs of the gap equation $\Delta_k = \pm 2E_k u_k v_k$. This translates by reversing the sign of $\Delta_k$ in Eq. (3). Thus, the interaction Hamiltonian is transformed to be

$$H_i = \frac{1}{2}\sum_{k,k',q} V_q \{\Psi^+_{k'-q}(t')I(k'-q,k')\Psi_{k'}(t')\}\{\Psi^+_{k+q}(t)I(k+q,k)\Psi_k(t)\}$$

$$-\sum_k \Psi_k^+(t)\{\chi_k I(k,k) - \Delta_k J(k,k)\}\Psi_k(t'), \qquad (22)$$

where $J(k_1,k_2) = M_{k_1}^{-1}\tau_1 M_{k_2}$. The second sum in Eq. (22) may be written as

$$\sum_k \Psi_k^+(t)\{\chi_k I(k,k) - \Delta_k J(k,k)\}\Psi_k(t')$$

$$= \sum_k \Psi_k^+(t)\frac{1}{E_k^2}\{\tau_3[\chi_k(\varepsilon_k^2 - \Delta_k^2) + 2\varepsilon_k\Delta_k^2] + \tau_1[2\varepsilon_k\Delta_k\chi_k - (\varepsilon_k^2 - \Delta_k^2)\Delta_k]\}\Psi_k(t').$$

$$(23)$$

The second sum is obtained by a field perturbation of the first sum in accordance with the Fock approximation. Thus, we simulate the obtainment of the Fock integral by applying the Wick's theorem on the first sum. To do so we insert the time ordering operator on the left side, fix **k**, assume $\mathbf{k'} = \mathbf{k} + \mathbf{q}$, and sum on **q**. Contracting $\Psi_{k'}(t')\Psi_{k'}^+(t)$ yields the Green's function $G_{0k'}(t-t')$ of Eq. (15). Let us calculate the product of the numerator of this function with the two interaction vertices, namely $\{I(k,k')[\omega'I + \varepsilon_{k'}\tau_3 + \Delta_{k'}\tau_1]I(k,k')\}$. It is given by

$$\{I(k,k')[\omega'I + \varepsilon_{k'}\tau_3 + \Delta_{k'}\tau_1]I(k,k')\} =$$

$$\frac{\tau_1}{E_k^2 E_{k'}^2}\{\Delta_{k'}[-(\varepsilon_k\varepsilon_{k'}-\Delta_k\Delta_{k'})^2+(\varepsilon_k\Delta_{k'}+\varepsilon_{k'}\Delta_k)^2]+2\varepsilon_{k'}(\varepsilon_k\varepsilon_{k'}-\Delta_k\Delta_{k'})(\varepsilon_k\Delta_{k'}+\varepsilon_{k'}\Delta_k)\}$$

$$+\frac{\tau_3}{E_k^2 E_{k'}^2}\{\varepsilon_{k'}[(\varepsilon_k\varepsilon_{k'}-\Delta_k\Delta_{k'})^2-(\varepsilon_k\Delta_{k'}+\varepsilon_{k'}\Delta_k)^2]+2\Delta_{k'}(\varepsilon_k\varepsilon_{k'}-\Delta_k\Delta_{k'})(\varepsilon_k\Delta_{k'}+\varepsilon_{k'}\Delta_k)\}$$

$$+\frac{I\omega}{E_k^2 E_{k'}^2}[(\varepsilon_k\varepsilon_{k'}-\Delta_k\Delta_{k'})^2+(\varepsilon_k\Delta_{k'}+\varepsilon_{k'}\Delta_k)^2]. \tag{24}$$

After some simple algebra we get

$$\{I(k,k')[\omega'I+\varepsilon_{k'}\tau_3+\Delta_{k'}\tau_1]I(k,k')\}=$$

$$\frac{1}{E_k^2}\{\tau_3[\varepsilon_{k'}(\varepsilon_k^2-\Delta_k^2)-\Delta_{k'}2\varepsilon_k\Delta_k]+\tau_1[\Delta_{k'}(\varepsilon_k^2-\Delta_k^2)+\varepsilon_{k'}2\varepsilon_k\Delta_k]+I\omega'\}. \tag{25}$$

Comparing Eqs. (23) and (25), and assuming that $\chi_k$, and $\Delta_k$ on the left hand sides of the integrals are related to $\varepsilon_{k'}$, and $\Delta_{k'}$ in the integrands, respectively, we get

$$\chi_k(\omega)=i\int\frac{d\omega'd^3k'}{(2\pi)^4}V(\omega-\omega',k-k')\frac{\varepsilon_{k'}}{\omega'^2-\varepsilon_{k'}^2-\Delta_{k'}^2}, \tag{26a}$$

$$\Delta_k(\omega)=-i\int\frac{d\omega'd^3k'}{(2\pi)^4}V(\omega-\omega',k-k')\frac{\Delta_{k'}}{\omega'^2-\varepsilon_{k'}^2-\Delta_{k'}^2}. \tag{26b}$$

Notice that by assuming frequency dependent potential in Eqs. (26) we have made a short-cut, of the otherwise long procedure, of introducing the phonon field and carrying out contractions of its operators, along with the electronic contractions, according to the Wick's theorem.

A superficial examination of Eqs. (26) might suggest that, despite the principal departure from the GN procedure, we have eventually come back to its results. This conclusion, though, is only partial. The quantities $\chi_k$ and $\Delta_k$ of Eqs. (26a) and (26b) apply only to the unperturbed Green's function. Thus, the results for the coherence

field-$\Psi_k$, converges to the results of the GN theory when restricting ourselves to the unperturbed propagators. This is not surprising since we have constructed our coherence field to be compatible with the unperturbed results of the GN theory. Besides, Gorkov obtained the same results from the basic equations for the normal and anomalous Greens' functions, without resorting to perturbation theory [3]. The situation becomes different when one wants to use perturbation theory to approximate beyond the first order Fock approximation of the unperturbed Hamiltonian. To attempt this we use the Dyson equation

$$G^{-1}(\omega,k) = G_0^{-1} - \Sigma_I I - \Sigma_3 \tau_3 - \Sigma_1 \tau_1, \tag{27}$$

where the $\Sigma$'s are the irreducible self-energies. The Greens' function of the full Hamiltonian is given by

$$G(\omega,k) = \frac{\omega Z I + \tilde{\varepsilon}_k \tau_3 + \varphi_k \tau_1}{\omega^2 Z^2 - \tilde{E}_k^2 + i\delta}, \tag{28}$$

with $\tilde{\varepsilon}_k = \varepsilon_k + \Sigma_3$, $\varphi_k = \Delta_k + \Sigma_1$, $\tilde{E}_k^2 = \tilde{\varepsilon}_k^2 + \varphi_k^2$, and $\Sigma_I(\omega) = \omega[1 - Z(\omega)]$. The calculation of the $\Sigma$'s is different than the calculation which led to Eqs. (26) in two respects: 1) The prefactors $[(\varepsilon_k^2 - \Delta_k^2)/E_k^2]$ and $[2\varepsilon_k \Delta_k / E_k^2]$ on the right hand side of Eq. (25) are not compensated. 2) The calculation of the $\Sigma$'s involves the fully dressed Greens' function G, rather than $G_0$. There is a common feature, though, and it is that the one body interaction of Eq. (22) applies to both cases, compensating partially the contribution of the first term. Thus, Eq. (23) is applicable, and the equivalent of Eq. (25) should be

$$\{I(k,k')[\omega' Z' I + \tilde{\varepsilon}_{k'} \tau_3 + \varphi_{k'} \tau_1] I(k,k')\} =$$

$$\frac{1}{E_k^2}\{\tau_3[\tilde{\varepsilon}_{k'}(\varepsilon_k^2 - \Delta_k^2) - \varphi_{k'} 2\varepsilon_k \Delta_k] + \tau_1[\varphi_{k'}(\varepsilon_k^2 - \Delta_k^2) + \tilde{\varepsilon}_{k'} 2\varepsilon_k \Delta_k] + IZ(\omega')\omega'\}. \tag{29}$$

The equations for the fully dressed irreducible self-energies then become

$$\Sigma_3(\omega,k) = \frac{i}{E_k^2} \int \frac{d\omega' d^3k'}{(2\pi)^4} V(\omega-\omega',k-k') \frac{(\varepsilon_k^2-\Delta_k^2)\tilde{\varepsilon}_{k'} - 2\varepsilon_k\Delta_k\varphi_{k'}}{Z'\omega'^2 - \tilde{\varepsilon}_{k'}^2 - \varphi_{k'}^2}$$

$$-\frac{1}{E_k^2}[\chi_k(\varepsilon_k^2-\Delta_k^2) + 2\varepsilon_k\Delta_k^2], \tag{30a}$$

$$\Sigma_1(\omega,k) = \frac{i}{E_k^2} \int \frac{d\omega' d^3k'}{(2\pi)^4} V(\omega-\omega',k-k') \frac{(\varepsilon_k^2-\Delta_k^2)\varphi_{k'} + 2\varepsilon_k\Delta_k\tilde{\varepsilon}_{k'}}{Z'\omega'^2 - \tilde{\varepsilon}_{k'}^2 - \varphi_{k'}^2}$$

$$-\frac{1}{E_k^2}[-\Delta_k(\varepsilon_k^2-\Delta_k^2) + 2\varepsilon_k\Delta_k\chi_k], \tag{30b}$$

$$\Sigma_I(\omega,k) = i\int \frac{d\omega' d^3k'}{(2\pi)^4} V(\omega-\omega',k-k') \frac{Z(\omega')\omega'}{Z'\omega'^2 - \tilde{\varepsilon}_{k'}^2 - \varphi_{k'}^2}. \tag{31}$$

Obviously, Eqs. (30) are different from their GN counterparts: 1) There is a mixing between the $\tau_1$ and the $\tau_3$ components. The two components of G contribute to each component of the self-energy. 2) There are the k-dependent coherence pre-factors $[(\varepsilon_k^2-\Delta_k^2)/E_k^2]$ and $[2\varepsilon_k\Delta_k/E_k^2]$, that weigh the two different components. 3) The self-energies- $\Sigma_1$ and $\Sigma_3$, are small corrections to be added to the Fock integrals $\Delta$ and $\chi$. This is a manifestation of the fact that the BV quasi-particles are the excitations of the almost accurate Hamiltonian, therefore, exerting only weak interactions. Note that Eqs. (30) reduce to the GN equations when $\Sigma_1$ and $\Sigma_3$ are set equal to zero, and $\chi_k$ and $\Delta_k$ are assumed to be related, respectively, to the $\tilde{\varepsilon}_{k'}$ and $\varphi_{k'}$ in the numerator of the integrand. Eq. (31) is essentially different from Eqs. (30), because the $I$ component of the self-energy cannot be incorporated within the time independent unperturbed Hamiltonian. It is obtained only by perturbation theory, and it is identical with the $I$ component of the GN theory. This is not surprising since superconductivity is known to have very little effect on the frequency renormalization.

Altogether, the corrections we have discussed so far to the GN theory are small in practice, despite the fundamental detour that we have undergone, with the coherence field. The main differences, which are the $\tau_1$ vertex and the coherence factors, are not effective in practice. However, so far we have discussed only Fock integrals. The allowance of Hartree integrals (for the pairing self-energy) makes a conceptually major difference. The Hartree integrals that contribute to the pairing self-energy are depicted in Fig. 1. Their sum is given by

$$\phi_H(k) = -iI_1(k,k)V(0,0)\int \frac{d^3k'd\omega'}{(2\pi)^4} Tr[I(k',k')G(\omega',k')\exp(i\delta\omega')], \quad (32)$$

$$= \frac{2\Delta_k \varepsilon_k}{E_k^2} NV(0,0), \quad (33)$$

where $I_1(k,k) = 2\Delta_k\varepsilon_k/E_k^2$, is the $\tau_1$ component of $I(k,k)$, and $N$ is the density of the electrons in the band. The result may be divided into two factors, the coherence factor $I_1(k,k)$, times the factor $NV(0,0)$. The latter is the Hartree potential in the normal state. Here it has been resulted from $Tr[I(k',k')G(\omega',k')]$ (which yields $\varepsilon_{k'}$ in the numerator of the integrand). Note that the quantities $\Delta_k$, and $E_k$ in the coherence factor $I_1$, are quantities of the unperturbed Hamiltonian. This coherence factor introduces a new feature to a component of the pairing self-energy, its anti-symmetry with respect to $\varepsilon_k$. Note also that whenever $\varphi_H$ is finite, it adds to the pairing self-energy, namely $\varphi_k = \Delta_k + \Sigma_1 + \varphi_H$.

In the next section we shall show that the factor $NV(0,0)$ vanishes in ordinary metals, because the static and long range interaction $V(0,0)$ is zero. This is so because, in the discussed limit, the Coulomb interaction is exactly compensated by the el-phonon-el interaction. However, the situation is not so in all cases. In the next section we conjecture that in HTSC, $\varphi_H$ might be finite and large.

## 4. CONCLUSION

We have demonstrated that there are two fields (and two more analogous fields) that are consistent with Wick's theorem, with the proper Hamiltonian and the proper unperturbed propagator. One is the well-known Nambu's field, and the other is the coherence field. We have shown that the coherence field yields the same results within the Fock approximation. The results of the higher order Fock approximation of Eqs. (30) are also very close to the results of the GN theory. In all these results the coherence factors, despite quite different interaction vertices, somehow sum up to yield the GN results.

An essential deviation from this tendency is the very existence of the Hartree self-energy, which exists in principle for the coherence field, but does not exist for the GN field. However, a closer examination shows that in all ordinary superconductors the Hartree pairing self-energy vanishes. The reason is the vanishing of $V(0,0)$, the static long range total interaction. We suppose that this feature holds in general for ordinary metals, but in the following we show it only for a simple mono-atomic model. For the bare el-phonon interaction g, we assume a model where the ions move as rigid bodies, and the electrons response adiabatically to their movements. Effects of inner deformations of the ions are incorporated in the dressed el-phonon interaction- $\bar{g}$ [10]. For such a model the squared bare el-phonon interaction is given by

$$g^2(q,\lambda) = \Omega_p^2 V_q \cos^2\theta_{q,\lambda} / 2\Omega_{q,\lambda}. \tag{34}$$

In Eq. (34), $V_q = 4\pi e^2 / q^2$, $\Omega_p$ is the ionic plasma frequency, $\Omega_{q,\lambda}$ is the frequency of the bare $\lambda$ phononic mode, and $\theta_{q,\lambda}$ is the angle between the polarization of the mode and the wave-vector **q**. The total interaction is the sum of the screened Coulomb interaction and the el-phonon-el interaction. The static total interaction is given by

$$V(0,\mathbf{q}) = \frac{V_q}{\varepsilon(0,q)} + \sum_\lambda \bar{g}^2(q,\lambda) D_{q,\lambda}(\omega=0). \tag{35}$$

In Eq. (35), $\varepsilon(\omega,q)$ is the dielectric function, $\bar{g}(q,\lambda) = g(q,\lambda)/\varepsilon(0,q)$ is the screened el-phonon matrix element, $\omega_{q,\lambda}$ is the screened phonon frequency, and $D_{q,\lambda}(\omega) = 2\Omega_{q,\lambda}/(\omega^2 - \omega_{q,\lambda}^2 + i\delta)$, is the phonon propagator. Inserting Eq. (34) into Eq. (35) yields

$$V(0,q) = \frac{V_q}{\varepsilon(0,q)}\{1 - \sum_\lambda \frac{\Omega_p^2 \cos^2\theta_{q,\lambda}}{\varepsilon(0,q)\omega_{q,\lambda}^2}\}. \tag{36}$$

In the sum over the three acoustic modes only the longitudinal mode contributes (because for the transversal modes $\cos\theta_{q,\lambda} = 0$). Moreover, for $q \to 0$, the "jellium" model in which $\omega_{q,L}^2 = \Omega_p^2/\varepsilon(0,q)$ applies. So we get

$$V(0,q) \to 0, \text{ for } q \to 0. \tag{37}$$

Although we have proved Eq. (37) only for a simple model, we assume its validity in more general metallic cases. Preliminary calculations show that it applies even for metals with optical modes. However, the key to its validity is the zero wavenumber. In cases where the Hartree diagram might be related to a finite wavenumber, the validity of Eq. (37) is no longer relevant.

The Hartree diagram corresponds to zero energy and momentum, because of the energy and momentum conservation along the successive parts of each term in the Feynman-Dyson equation. The two unperturbed propagators on both sides of the Hartree diagram must have the same energy and momentum parameters, resulting in zero energy and momentum for the self-energy itself. In HTSC, however, there are some theoretical implications [1,2], and some experimental evidences [11-13], that momentum conservation, in directions normal to nesting surfaces, may be violated by twice the Fermi momentum. This might open the possibility of having a Hartree diagram that carries such a momentum. If so, such a self-energy could be exceptionally large. However, more analysis of the present type is still needed to establish this possibility. So far, it is only a conjecture.

The present analysis proved that there are two different fields that are consistent with the application of Wick's theorem, with the Hamiltonian and the unperturbed propagator. What is the single field that is appropriate for superconductivity? We have argued that, for all conventional superconductors, the two fields are consistent with experimental results. The main difference is the admittance of the Hartree pairing self-energy for the coherence field, and its forbiddance for the GN field. But could this be considered an essential difference when the Hartree diagram vanishes anyway? A negative answer to this question makes the dilemma irrelevant. For HTSC with pseudogaps in the normal state, we conjectured that the violation of momentum conservation, might result in a finite Hartree pairing self-energy. However, since this analysis has not yet been carried out, it is still premature to deal with the question of selecting the right field.

# FIGURE CAPTIONS

Fig. 1

The two Hartree diagrams that contribute to the Hartree pairing self-energy. They are different only in the components of the vertex and the bubble propagator,

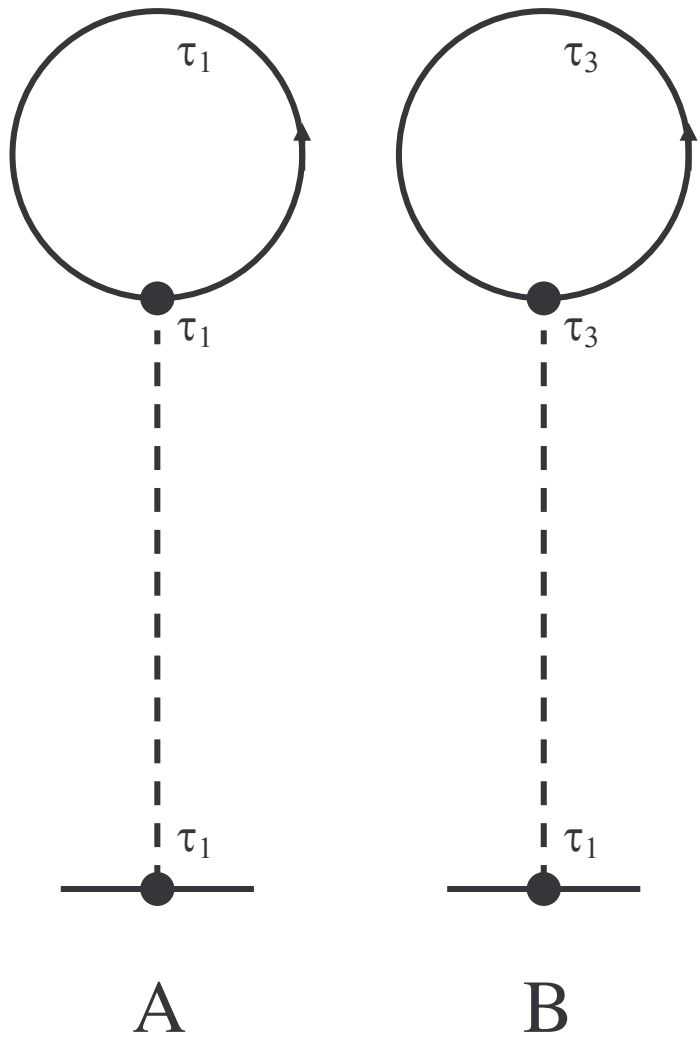